\documentclass{ws-procs975x65}

\begin{document}

\title{QUANTUM LIOUVILLE THEORY WITH HEAVY CHARGES
\footnote{Presented by Pietro Menotti}}

\author{PIETRO MENOTTI}

\address{Dipartimento di Fisica, Universit\`a di Pisa and \\
INFN Sezione di Pisa\\
\email{menotti@df.unipi.it}}

\author{ERIK TONNI}
\address{Scuola Normale Superiore, Pisa and\\
INFN, Sezione di Pisa\\ 
\email{e.tonni@sns.it}}


\begin{abstract}

We develop a general technique for solving the Riemann-Hilbert problem
in presence of a number of "heavy charges" and a small one thus
providing the exact Green functions of Liouville theory for various
non trivial backgrounds. The non invariant regularization suggested
by Zamolodchikov and Zamolodchikov gives the correct quantum
dimensions; this is shown to one loop in the sphere topology and for
boundary Liouville theory and to all loop on the pseudosphere. The
method is also applied to give perturbative checks of the one point
functions derived in the bootstrap approach by Fateev Zamolodchikov and
Zamolodchikov in boundary Liouville theory
and by Zamolodchikov and Zamolodchikov on the pseudosphere, obtaining
complete agreement.
\end{abstract}

\bodymatter

\section{Introduction}\label{intro}

Liouville theory has attracted a lot of interest as an example of
quantum conformal field theory \cite{CT} and for its applications to
model string theory and to brane theory. Remarkable results have been
obtained within the bootstrap approach \cite{bootstrap,ZZpseudo}, which
starting from some assumptions provides exact results for a few
interesting correlation functions.

Here we address the problem to recover the conformal quantum Liouville
field theory from the functional integral procedure understood
in the usual sense in which one starts from a stable background and
then one integrates over the quantum fluctuations.  As it is well
known, a quantum field theory is specified not only by an action but
also by a regularization and renormalization procedure. 

Both on the sphere topology formulated on the Riemann sphere,
on the pseudosphere and obviously in the 
conformal boundary case, the Liouville action has to be supplemented by
boundary terms. For definiteness we shall illustrate here the conformal
boundary case. The action in presence of sources is given by
\begin{eqnarray}\label{boundaction}
\hspace{0cm} S_{\,\Gamma,\,N}[ \,\phi\,] & = &
 \lim_{\varepsilon_n\,\rightarrow\, 0}\,
 \Bigg\{ \int_{\Gamma_{\varepsilon}}
\left[ \,\frac{1}{\pi} \,\partial_\zeta \phi
\,\partial_{\bar{\zeta}}\phi+\mu\, e^{2b\phi}\,\right] d ^2 \zeta
\,+\, \oint_{\partial\Gamma} \left[ \,\frac{Q\,k}{2\pi}\,\phi
+\mu_{\scriptscriptstyle\hspace{-.05cm}B}\,
e^{b\phi}\,\right] d\lambda\,  \\
  &  & \hspace{1.4cm}\rule{0pt}{.9cm}
   -\,\frac{1}{2\pi i}\;\sum_{n=1} ^N
   \alpha_n\oint_{\partial\gamma_n} \phi
\left( \, \frac{d \zeta}{\rule{0pt}{.4cm}\zeta-\zeta_n}- \frac{d
\bar{\zeta}}{\rule{0pt}{.4cm}\bar{\zeta}-\bar{\zeta}_n}\, \right)-
\sum_{n=1} ^N \alpha_n^2 \log \varepsilon_n^2 \; \Bigg\} \nonumber
\end{eqnarray}
where the
integration domain $\Gamma_{\varepsilon}=\Gamma\hspace{-.07cm}
\setminus \bigcup_{n=1}^N \gamma_{n}$ is obtained by removing $N$
infinitesimal disks $\gamma_{n}=\{|\zeta-\zeta_n|<\varepsilon_n\}$
from the simply connected domain $\Gamma$ and $\phi\approx -\alpha_n
\log|\zeta-\zeta_n|^2$ for $\zeta\rightarrow \zeta_n$.
$Q=1/b + b$ and $k$ is the extrinsic curvature of the boundary
$\partial\Gamma$, defined as
\begin{equation}\label{k definition}
k\,=\,\frac{1}{2i}\;\frac{d}{d\lambda}
\left(\log\frac{d\zeta}{d\lambda}\,-\,
\log\frac{d\bar{\zeta}}{d\lambda}\,\right)
 \hspace{1,1cm},\hspace{1,1cm} \zeta(\lambda)\in\partial\Gamma
\end{equation} where $\lambda$ is the parametric boundary length, i.e.
$d\lambda=\sqrt{\rule{0pt}{.386cm}d\zeta d\bar{\zeta}}$.  It is
possible to write action (\ref{boundaction}) as the sum of a classical
part and quantum action. One notices
that due to $Q\neq 1/b$ the above written action is not exactly
invariant under conformal transformations. In \cite{MT2,MV}
it was found that if one starts from $Q=1/b$ and adopts an invariant
regularization procedure one does not reach a theory invariant under 
the full
conformal group. This is similar to the result of \cite{DFJ} . The
reason is that in such an approach the cosmological term $e^{2b\phi}$
acquires weight $(1-b^2,1-b^2)$ instead of \cite{MT2} $(1,1)$  as
required by the full infinite dimensional conformal invariance.

The regularization suggested at the perturbative level in
\cite{ZZpseudo} in the case of the pseudosphere provides the vertex
functions with the correct quantum dimensions \cite{CT} at the first
perturbative order $\Delta_{\alpha}=\alpha(1/b+b-\alpha)$. 
In \cite{MT3-4} is was explicitly proven that such
a result stays unchanged to all orders perturbation theory. In
particular the weight of the cosmological term becomes $(1,1)$ as
required by the invariance under local conformal transformations.
These
calculations correspond to a double perturbative expansion in the
coupling constant and in the charge of the vertex function. 

Here we use a more powerful approach which allows to resum infinite
classes of graphs \cite{MV}~. 
We start from the background generated by finite charges,
i.e. ``heavy charges'' in the terminology of \cite{ZZsphere}~. This
means that we consider the vertex operators $V_{\alpha_n}(z_n) =
e^{2\alpha_n \phi(z_n)}$ with $\alpha_n=\eta_n/b$ and $\eta_n$ fixed
in the semiclassical limit $b \rightarrow 0$. This has the remarkable
advantage to give the resummation of infinite classes of usual
perturbative graphs. In order to do that however one needs
the exact Green function on a non trivial background.

In the case of a single heavy charge, by solving a Riemann-Hilbert
problem in presence of the given heavy charge and an infinitesimal
one we are able to compute such exact Green function on such a
background in closed form in terms of incomplete Beta functions and
such a Green function is used to develop the subsequent perturbative
expansion in the coupling constant $b$.

After such a result is accomplished one is faced with the non trivial
task of computing a functional integral constrained by the boundary
conditions imposed by action (\ref{boundaction}).

The background generated by a single
charge is stable only in presence of a negative value of $b^2
\mu_B$. 
We compute the Green function on such a background
satisfying the correct conformally invariant boundary conditions and
such a Green function is regularized at coincident points by simply
subtracting the logarithmic divergence.
For the sphere and conformal boundary case one obtains the correct
quantum dimensions to one loop in such 
background improved perturbation theory.
The presence of a negative boundary cosmological constant imposes to
work with the fixed boundary length $l$ constraint and to compare our
results with the ones given in \cite{FZZ} also the fixed area $A$
constraint is introduced. It is possible to factorize the functional
integral in a term resulting from the boundary length and area
constraints and an unconstrained
functional on functions satisfying the correct conformal invariant
boundary condition. We compute such functional integral through the
technique of varying the charges and the invariant ratio $A/l^2$.
The one loop result on the one source background obtained in this way
is \cite{MT5}
\begin{equation}\label{ourresult}
Z(\eta;A,l\hspace{.04cm})\,=\,e^{-S_{0}(\eta;A,l)/b^2}\;
\frac{\beta}{8\pi^2} \;\frac{l}{b^2 A}\;
\frac{e^{2\eta\gamma_{\,\scriptscriptstyle\hspace{-.05cm}E}}\,
\Gamma(2\eta)}{\sqrt{1-2\eta}} \,\big(1+O(b^2)\big) \end{equation}
where $S_{0}(\eta;A,l)/b^2$ is the classical action without the bulk
and boundary cosmological terms, computed on the
one source background. Eq.(\ref{ourresult}) agrees with the expansion of the
fixed area and boundary length one 
point function derived through the bootstrap method in \cite{FZZ} and
for which there was up to now no perturbative check.

Applying similar techniques in the pseudosphere case one obtains
for the one point function
\begin{equation}\label{onepointpseudo}
\langle \, V_{\eta/b}(0) \,\rangle = e^{-S_{cl}(\eta)/b^2}
\;\frac{e^{2\eta\gamma_{\,\scriptscriptstyle\hspace{-.05cm}E}}}
{\Gamma(1-2\eta)\,(1-2\eta)^{3/2}}
 \;\big(\,1+O(b^2)\,\big)\;
\end{equation}
where $S_{cl}$ is the full classical action
and the one loop expression  for the two point function due to a
finite charge and an infinitesimal one. Eq.(\ref{onepointpseudo})
agrees with the expansion of the bootstrap result, while the
expression for the two point function is consistent with the
existing results of the standard perturbation approach and agrees with
the exact two point function when one vertex is given by the degenerate
field $V_{-1/ (2b)}$. 
On the other hand adopting the invariant regularization for the Green
function at coincident points one finds an expression which disagrees
with  the degenerate two point function
$
\langle V_{-1/(2b)}(x) V_{\varepsilon/b}(y)\rangle
$
on the pseudosphere.

\vfill

\end{document}